\documentstyle[11pt,newpasp,twoside,epsf]{article}
\begin{document}

\title{Conference Summary: Mapping the Hidden Universe}
\author{R. Brent Tully}
\affil{Institute for Astronomy, University of Hawaii, Honolulu, HI 96822, U.S.A.}

\begin{abstract}
Two curiously connected topics provided a novel focus for this meeting
in Guanajuato, Mexico, 22--29, February, 2000.  Two days were devoted to 
discussions of galaxy surveys deep into the Galactic plane and to modeling
of the distribution of matter that makes use of quasi-full-sky coverage.
Then the meeting turned its attention for four further days to a look at
the Universe from surveys based on HI selection.  How distorted is the
world view of the optical chauvinist?

\end{abstract}

\section{The Universe Behind the Milky Way}

Amazing!  The gap in maps of the distribution of known galaxies caused by
foreground obscuration is almost disappearing.  We currently have coverage
of about 90\% of the sky.

Why do we care if we miss part of the sky?  All the kinds of galaxies that 
make up the Universe
are probably well represented at high latitudes, so why struggle to study
extragalactic targets dimmed by intervening material?  However, for some 
purposes it is not the
individual galaxies that interest us so much as it is their {\it distribution}.
We cannot
study structures formed by galaxies that are larger than the bounding box of 
our surveys.  Boundaries are
created either by distance, the obscuration of the Milky Way, or observing
exigencies.  The obscuration boundary is nasty because it is often soft and
it splits samples where we appreciate it the least; nearby where otherwise
our knowledge is the best.

More fundamental than the distribution of light is the distribution of mass.
We know from the microwave
background dipole anisotropy that we are being tugged at 600 km/s toward
a location in the sky uncomfortably close to the Galactic plane.
Detailed mapping of the peculiar velocities of galaxies can provide a
description of the underlying distribution of matter.  Comparison can be
made between the distribution of luminous objects and the inferred 
distribution of matter to gain a better understanding of the nature of
the mass content of the Universe.  Our ability to separate peculiar velocities
from the Hubble flow fails linearly with distance so dynamical tests are best
done in the nearby volume.  Modeling could be seriously compromised if a
massive veiled neighbor is missed.

Our host Ren\'ee Kraan-Korteweg showed beautiful maps of the current
status of observations at low Galactic latitudes.  Presumably examples of
the maps have found their way into this book, which henceforth will be
referenced as `proceedings'.  The new information that is contributing to
these improved maps is coming from a wide range of techniques, which is
necessary because no one technique is fully adequate.  We heard about:

\noindent$\bullet$
Spectroscopic surveys of optically selected galaxies (Weinberger et al.,
proceedings; Fairall \& Kraan-Korteweg, proceedings; Roman et al.,
proceedings; Pagani et al., proceedings; Wakamatsu et al., proceedings;
Woudt et al., proceedings; Pantoja, proceedings).

\noindent$\bullet$
Near or far infrared selection with either optical or HI spectroscopic 
follow up
(Huchra et al., proceedings; Schr\"oder et al., proceedings; Ragaigne et al.,
proceedings; Saunders et al., proceedings; Vauglin et al., proceedings;
Nakanishi et al., proceedings).

\noindent$\bullet$
X-ray selection with spectroscopic follow up (Ebeling et al., proceedings;
B\"ohringer, proceedings).

\noindent$\bullet$
Radio continuum selection (Green et al., proceedings).

\noindent$\bullet$
Blind HI surveys (Henning, proceedings; Staveley-Smith, proceedings).

The need for this multi-pronged approach is evident enough.  Infrared 
radiation penetrates the zone of obscuration although there can be overwhelming
confusion at the lowest Galactic latitudes.  Spirals galaxies are strong
sources of far infrared emission and contain HI that can be observed for a
redshift confirmation.  Early type galaxies are rarely detected in the far
infrared but these high surface brightness objects can be found with near
infrared surveys and redshifts can be determined with low dispersion
spectrographs on large optical telescopes if obscuration is not extreme.  
The dense environments of
clusters of early type galaxies are found with the X-ray surveys.  These
clusters mark the deepest gravitational potential wells.  Radio continuum
surveys detect a wide range of galaxy types and can be cross-correlated
with infrared surveys to winnow candidate lists at low latitudes.
Blind HI surveys provides almost the only means of detecting low surface
brightness late type galaxies and can provide unhindered detection of
spiral galaxies at even the lowest Galactic latitudes.

Hence we have the tools to greatly diminish the restrictions caused by
our vantage point in the Milky Way and the surveys are well under way.
The current observational status is still a bit of a hodge-podge, varying
with sector along the band of the Galactic plane.  Roughly speaking,
there is good completion now to $\mid b \mid \sim 5^{\circ}$.  In some
directions toward the Galacic anti-center, there is good information right
down to the Galactic equator.

Where do things stand?
I will stick my neck out and claim that we are probably 
{\it not} missing a dynamically important individual galaxy from our current
inventory.  It is to be remarked that the two galaxies outside the Local 
Group with the largest influence on us {\it are} at very low Galactic
latitudes; IC~342 and Maffei~I.  Their very names tell you that they are
obscured (ie, not Messier or NGC objects).  For an individual galaxy to
be dynamically important it should be within $\sim 5$~Mpc and have a 
luminosity greater than $\sim 10^{10}~L_{\odot}$.  It is unlikely that such
an object would evade the filters described above.

On the other hand, our current maps of the collectivity of galaxies into
large scale structure are incomplete.
From continuity of observed structure, we can make reasonable guesses about
where future pickings will be most abundant.  Fortuitously, our galaxy is
face-on to the biggest nearby structures, those linking to the Virgo and Fornax
clusters, and edge-on to rather empty regions.  Several filamentary strands
penetrate this low density space, including one that contains our Galaxy.  
It is not
until a redshift of about 3000~km/s that we run into serious structure in
the zone of obscuration in the direction of the `Great Attractor'.
Important structures run from the Centaurus Cluster through the plane to
the Norma Cluster and from the Hydra Cluster off, ultimately, to the Perseus
Cluster region.  The low latitude  Norma Cluster is rich and appears to be 
a particularly important nodal point (Woudt, proceedings).  Behind this
region there are hints of even bigger structures, possibly linking the
huge Shapley Concentration and Horologium-Reticulum cluster complexes
(Ebeling et al., proceedings).

These observations are leading to a better understanding of structure
formation and evolution.  Contributions to this effort heard/seen at this 
conference include D'Mellow et al. and Valentine et al. (proceedings), 
Hoffman (proceedings), and Zaroubi (proceedings).  Interestingly, there are
strong hints that roughly half the motion of the dipole anisotropy is
generated beyond 6,000~km/s but probably within 15,000~km/s.  Much of the
action is at low Galactic latitudes.

Credit for the improved observational situation can be given
to Ren\'ee Kraan-Korteweg, not only for her contributions to the observational
programs but also for hosting two seminal meetings on the topic.  The first
meeting in Paris (Balkowski \& Kraan-Korteweg 1994) focused attention on the 
problem and
this second meeting has provided a stimulus for a lot of mid-course activity.
A basic picture is beginning to jell.  We wonder if Ren\'ee can find an
equally spectacular third location for the rap-up conference a few years from 
now.

\section{The Universe in HI}

On this subject, I have an attitude.  The important question is whether there
are optically undistinguished objects that can be detected in HI and that
constitute an important part of the inventory of objects in the Universe.

Let me first explain the source of my attitude.  Two decades ago in the
course of an HI survey (Fisher \& Tully 1981$a$), I joined the ranks of only a
few people and inspected {\it every} sky atlas print in
considerable detail (about a half hour per print).  There were many
objects on the prints with such low surface brightnesses that they could 
barely be seen, yet 
these objects were detected with high efficiency with subsequent HI pointings.
It seemed reasonable to us that there would be objects with yet lower
surface brightnesses, objects that would be invisible against the sky at
optical bands but that would have detectable HI.  We looked hard but
could not find them (Fisher \& Tully 1981$b$).  Every pointed observation
in our program could be inspected for `second' HI signals, signals due to 
something besides the
primary target.  Every on-target pointing was accompanied by an off-target
pointing at a random position that could be inspected.  In addition, we 
surveyed blank sky at over a thousand beams in selected regions like the 
equatorial plane of the Local Supercluster where dwarf galaxies might be 
expected.

We recorded serendipitous detections.  {\it In a majority of cases}, the
blind detections lie at the positions of obvious spiral galaxies.
{\it In almost all other cases}, the detections lie at the positions of 
low surface brightness galaxies that were discernible on the sky surveys
(also Huchtmeier, proceedings).
{\it In the rare remaining cases with significant HI signals}, mapping of
the HI sources revealed that the HI was associated with extended structures
around individual galaxies or in groups and could be interpreted as 
tidal in origin.  No invisible galaxies were found.

The speculation about the possible existence of a cosmologically important
class of ultra low surface brightness galaxies was already in the air 
(Disney 1974) 
but seemed to us to be discounted by the observations mentioned above.
Still, maybe those earlier HI observations were just
not sensitive enough and maybe there were crouching giants below our flux
limits.  Since those early days, the much more sensitive Arecibo Telescope 
has observed
tens of thousands of sight lines (Haynes et al. 1997 and others).  
Yes, there have been a couple of curious situations reported (eg, Giovanelli
et al. 1991).
But maybe the interesting thing is the {\it lack} of such reports.
At this conference we have begun to hear results from the systematic survey
with the Parkes Telescope multibeam system (Webster et al., proceedings;
Waugh et al., proceedings; Drinkwater et al., proceedings;
also Arecibo dual-beam: Rosenberg \& Schneider and Schneider \&
Rosenberg, proceedings).  These 
surveys are much to be appreciated because of their systematic nature.
It is my impression that they are confirming what some of us have
long suspected.

Aha, attitude.  These new studies tend to accent the fact that {\it they 
are making blind HI detections}.  The talks cited above announced faint end
HI mass function indices $-1.3 < \alpha < -1.7$.  Caution however: in samples 
assembled from the field (Webster, Schneider),
the mass function is constructed with contributions at the high mass end
from {\it distant} giants and contributions at the low mass end from 
{\it nearby} dwarfs.
How are these separate contributions to be normalized?  We know we live in
an overdense region locally.  And the statistics at
the low mass end are, frankly, terrible.  Verheijen et al. (proceedings)
find $\alpha = -1.1$ in the better controlled, volume limited environment
of the Ursa Major Cluster.  Van Woerden et al. (proceedings) and Zwaan \&
Briggs (proceedings) discuss High Velocity Clouds and discount the
suggestion that they are a significant extragalactic component.

Is the glass half full or half empty?  The standard plots of the HI mass
function show numbers of galaxies per mass bin and the `half full'
viewpoint stresses that numbers tend to rise
toward low masses (albeit minimally according to Verheijen et al.).
My Figure~1 combines the data in Fig.~2 of Verheijen et al.
with data from the Parkes multi-beam survey of the Centaurus Group
(Banks et al. 1999) and optically identified members of the Local Group.
However, my `half empty' plot converts from number of objects per bin to 
a histogram of HI mass
per bin.  {\it Overwhelmingly, the global HI mass resides in a small number of
big galaxies.} 
Dwarf galaxies can have a lot of gas for their light and are reasonably 
numerous but still contribute only a tiny fraction of the neutral Hydrogen
inventory in the zero redshift Universe.  Driver (proceedings) makes a
related point about luminosities and total mass.
This highly probable conclusion about the location of HI reservoirs
explains
why a majority of blind HI detections are associated with big, visually
obvious galaxies.
Croutching giants are negligible in number.  These conclusions are not
new.  Zwaan et al. (1997) have said basically the same thing.

\begin{figure}
\plotone{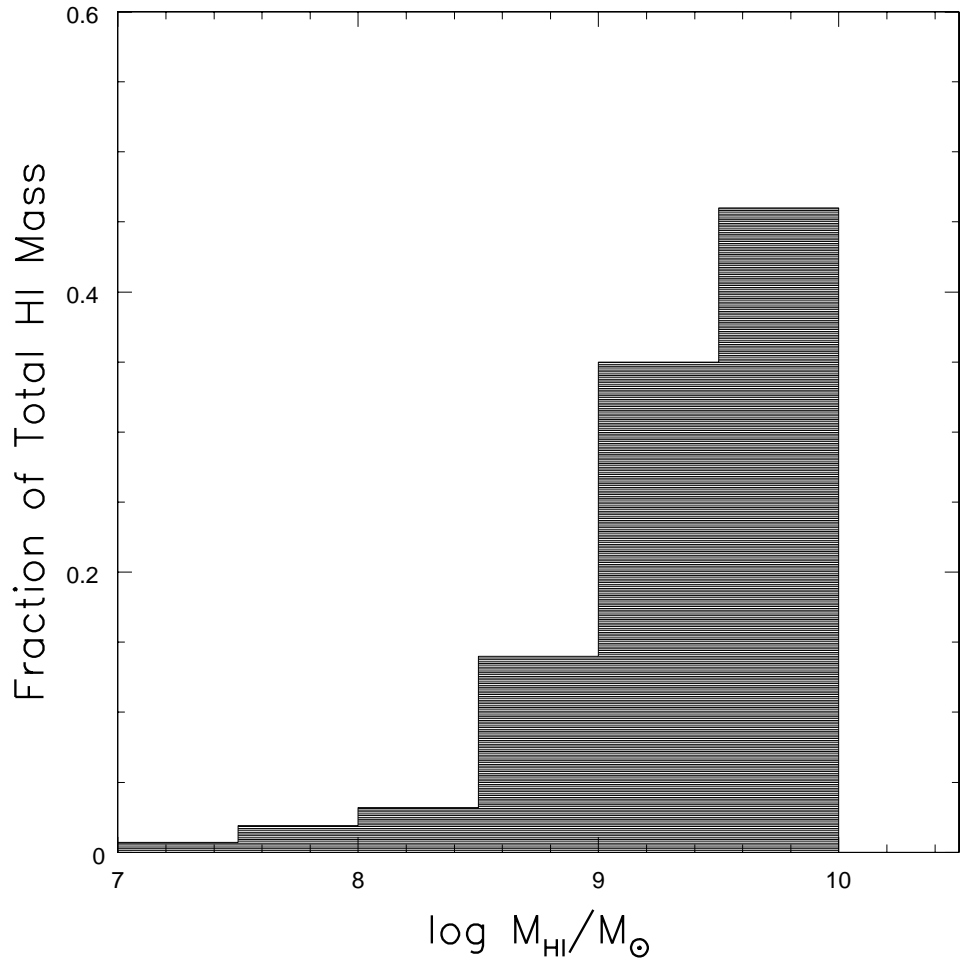}
\caption{Fraction of HI mass in logarithmic mass intervals.  Three samples 
are combined: Ursa Major Cluster, Centaurus Group, and Local Group.  All 3
samples are substantially complete above log~$M_{HI}/M_{\odot}=7$.  Most
of the HI is in systems with log~$M_{HI}/M_{\odot}=9.3\pm0.5$ }
\end{figure}

If the glass is half empty rather than half full, that is an interesting
result!
We are left to conclude that there is a lower density cutoff for HI systems.
Probably gas clouds below some threshold are ionized by the metagalactic
radiation field.  HI systems with densities above the cutoff might form stars
in the way outlined by Legrand (proceedings); also see Young (proceedings).
The star formation process is sufficiently well constrained over time that
the HI density threshold translates into a surface density of star light
that, by fluke, barely exceeds the blue light sky survey detection 
thresholds.  

To sum up:

\noindent$\bullet$
{\it There are NO dynamically important components of the $z=0$ Universe
that are detectable in HI but not in the optical.}  Still, classes of objects
like Low Surface Brightness galaxies and Blue Compact Dwarfs can be found 
with fewer observational biases in HI.  Of course, galaxies obscured by the
Milky Way might only be found by HI.

\noindent$\bullet$
{\it Most of the HI mass in the $z=0$ Universe is locked up in galaxies with
$8.5 < {\rm log} M_{HI}/M_{\odot} < 10$}.  There are not enough low mass
galaxies to make a major contribution to the global sum of HI.  At the high
mass end, evidently large HI reservoirs lead to highly efficient star 
formation and rapid depletion of the reservoirs.  Hence, the histogram of
HI mass per unit mass is more peaked than the equivalent histograms of 
light or total mass.

\noindent$\bullet$
{\it There must be an HI surface density cutoff at 
log~$n_H \sim 19.5$~atoms/cm$^2$.}  Neutral Hydrogen clouds can be 
maintained above this density and must form stars at a rate sufficient to
produce a surface brightness detectable at optical bands.  Below this
threshold, gas clouds in the $z=0$ Universe must be ionized.

\noindent$\bullet$
The conference has been most informative and entertaining, Guanajuato is
a wonderful town and site for the meeting, and hurray to the organizers,
Pat Henning, Heinz Andernach, all the people behind the scene, and,
especially, Ren\'ee Kraan-Korteweg.

\end{document}